\documentclass[twocolumn,showpacs,preprintnumbers,amsmath,amssymb]{revtex4}

\usepackage{graphicx}
\usepackage{dcolumn}
\usepackage{bm}

\begin{document}

\title{Cellular structure in system of interacting particles.}

\author{Bohdan Lev}

\affiliation{Bogolyubov Institute of Theoretical Physics, NAS,
Ukraine, Meytrologichna 14-b, Kyiv 03143, Ukraine.}

\date{\today}

\begin{abstract}
The general description of formation the cellular structure in the
system of interacting particles is proposed. Interactions between
particles are presumably well-understood and the phase transition
in which can be studied in the scale of particle resolution. We
presented analytical results of possible cellular structures for
suspension of colloidal particles, in system particles immersed
in liquid crystal and gravitational system. We have shown that 
cellular structure formation can occur in system of interacting 
particles for realistic values of temperature and particles concentration.
\end{abstract}


 \maketitle

At the same time the resulting physical properties in the system
of interacting particles have been the subject of much active
research. The phase separation and ordering of many particles
system and their resulting properties excite curiosity as
fundamental physics as large number of viable application. For
description the behavior the many particles system with different
character of interaction we must develop approach which taking
into account a spatial nonhomogeneous distribution of particle i.e
cluster formation or cellular structure and more ordering other
structures. Basically, statistical description of many particles
systems concern homogeneous states. The most interesting and
pressing problem in condense matter physic is the study of phase
transition with spatial nonhomogeneous distribution of particle-
cluster formation. This structure was observed in usual colloids
\cite{gas},\cite{low} in the system particle which introduced in
liquid crystal \cite{and1},\cite{and2}, \cite{chen} and even in distribution
galactic system \cite{sas}. As evident from this examples,
formation cellular structures in the system of interacting
particle have been large-scale demonstration for different
physical situation. Cellular structure in system particle was
dramatic change the properties of condense matter. Depending 
of the size of particle was observed the formation of both 
reversible and irreversible clustering. During the past decade, 
colloidal system have been used as model system to understand 
phenomena the two dimensional melting transition. In usual colloids
was observed the crystal like ordering, was observed hexagonal
crystal structure \cite{nych}. Experimentally, it is observed that the initial stage 
of the cellular structure originates from region with lowest local 
density of particles. In all case exist the elastic interaction 
leads to spatially inhomogeneous distribution of such particles with 
formation of regions of pure from particles. The energy interaction in system 
particle have different character and value. Minimum of free 
energy realizes cellular structure and leads to stabilization 
the regions of pure elastic media, and the particles themselves 
are placed at the boundaries of those regions. To describe the 
conditions of formation and properties of these particle structures,
one should take into account both aspects of the particle interaction. 
At first we shall give representation of the general approach in 
the description of behavior in system of cooperating particles. 
Let's show as in this system can be realized of a first order phase 
transition with creation of spatially non-uniform distribution of  
particles. The formation of different structures depends upon 
the initial concentration of particles. The observation of optimum 
cluster size has important consequences for the formation cellular 
structure. The system will try minimize the energy by breaking it apart,
and out of many ways it can be done, the system chooses the nucleation 
and further formation of cellular structures. 

We would like show an investigating phenomena that occurs in colloidal 
system the formation of colloidal voids and cluster structures. 
Theoretical investigation of colloidal system focused on thermodynamic 
treatment or on general arguments. The essential point which is neglected 
any at these  arguments is that the colloidal particles are interacting
via their direct forces and indirect interaction mediated by free
energy of solvent phases. We propose a general description which
include effect contains parameters whose relation to formation of
different structures in the system of interacting particles.
Basically it consider very know Landau approach for first order
phase transition coupled to the colloidal contribution and
combined with the direct interaction between the colloidal
particles. An analysis of this general model can predict a rich scenario
of solvent induced colloidal phase separation with formation of
cluster and cellular structures. 

The physical properties depend on the structure of on individual
particle. However practical exist within spatial nonuniform
distribution of particles, and so inter particles interactions
will be also influence the physical behavior of the this
distribution and also aggregation process by within the spatial
distribution formed. Characteristic of aggregation effect very
difficult, although there have been attempt to measure fractal
dimensional form of distribution particles. The presentation and
develop method for system of interacting particle, several
interesting feature have become evident. The method must describe 
to observations on similar system with different nature of interaction 
between particles which formed systems. Based on general approach we can 
estimate formation cellular structure in different system.

First from all we consider system of not interacting particles.
Instead of investigate the nonuniform distribution of particle we
describe possibility formation the cavity where not particles.
Empty cavity or pore in distribution particle is very simple find
and measurement. In the case contiguous distribution of particles
with concentration $c$ the size of pore can determine as distance
to nearest particles. We consider formation only cavity of
spherical form. The form of cavity not play principal role and
other form of  pore we take into account after more precise the
all process formation the cavity.  For determination of
probability of formation the cavity without particle used the
Saslaw approach \cite{sas} which can describe the distribution of
galactic. The probability depended from distribution of nearest
particles. For different point of space the probability $p(r)dr$
of disclose the particle on the distance between $r$ and $r+dr$
is equal the probability nonoccurrence of particles in the area
with size $r$ which must multiply on the probability to find the
particle on the distance $r$. This relation have the following
mathematical form:
\begin{equation}
p(r)dr=\left\{1-\int^{r}_{o}p(r')dr'\right\}4\pi c r^{2}dr
\end{equation}
This equation can rewrite in equivalent form:
\begin{equation}
\frac{d}{dr}\left\{\frac{p(r)}{4\pi c
r^{2}}\right\}=-p(r)=-\left\{\frac{p(r)}{4\pi c
r^{2}}\right\}4\pi c r^{2}
\end{equation}
which have solution
\begin{equation}
p(r)= 4\pi c r^{2} exp (-\frac{4\pi c r^{2}}{3})
\end{equation}
The probability $p(r)dr$ of disclose the particle on the distance
between $r$ and $r+dr$ is equal the probability nonoccurrence of
particles in volume $V$  multiply on the probability $cdV$to find
the particle in the volume between $V$ and $V+dV$: $p(r)dr=cPdV$.
From this relation as result we can obtain the probability
disclose the cavity without particle with volume $V$ in the form
\begin{equation}
P(V)= exp (-cV)=xp(-N)
\end{equation}
for Poison distribution. The main size of area without particle
can estimate from relation $P(V)\simeq \frac{1}{2}$. Obvious, that
in system of not interacting particle we can obtain only
nonuniform distribution of particles, but we have non zero the
probability disclose the cavity without particle with volume $V$.

Now, we can find analytical result for the probability disclose
the cavity without particle with random selection the volume $V$
in the case of system interacting particles. The probability the
presence in system interaction particles in ground canonical
assemble $N$ particles in volume $V$ at temperature $T$ can be
write in the standard form \cite{huang}:
\begin{equation}
W(N)=exp\left\{\beta \mu N-\beta F \right\}
\end{equation}
where $\beta=\frac{1}{kT}$ inverse temperature,$\mu$ chemical
potential, and $F(N,V,T)$ is free energy which can calculate with
the help of canonical assemble. With this probability to find the
mean value of derived magnitude $exp(qN)$ which must depend from 
number of particles
\begin{eqnarray}
&& \left\langle
exp(qN)\right\rangle=\sum_{N}exp(qN)W(N)=\nonumber\\
&&G^{-1}\sum_{N}exp\left\{(\beta
\mu+q) N-\beta F \right\}\nonumber\\
&&\equiv exp\left\{\Psi(\beta \mu+q)-\Psi(\beta \mu+q)\right\}
\end{eqnarray}
where we used the very know relation between statistical sums of
ground canonical assemble $G$ and thermodynamic function:
\begin{equation}
ln G(\mu,V,T)=\frac{PV}{kT}\equiv \Psi(\mu,V,T)
\end{equation}
The previous relation can rewrite in other form:
\begin{eqnarray}
&&\sum_{N}z^{N}W(N)=exp\left\{\Psi(z exp(\beta
\mu))-\Psi(exp(\beta \mu)\right\}=\nonumber\\
&& exp(-\Psi)exp(\Psi(z exp(\beta \mu)))
\end{eqnarray}
Now we can expansion in series at $z$ the right part this
relation and to compare the equal power of $z$ in both part
equation we obtain
\begin{equation}
W(N)=exp(-\Psi)\frac{exp(\beta \mu)}{N!}(exp \Psi)^{N}_{0}
\end{equation}
where
\begin{equation}
(exp \Psi)^{N}_{0}=\left[(\frac{d}{d(z exp(\beta \mu ))})^{N}exp
\Psi(z exp(\beta \mu))\right]
\end{equation}
when $z exp(\beta \mu=0$. After this operation we can obtain the
probability find the empty cavity in distribution of interacting
particles. This probability can write in the form:
\begin{equation}
W(0)=exp(-\frac{PV}{kT})
\end{equation}
If we know the equation of states we can estimate in general case
the probability existence the cavity without particle in the case
of system interacting particles. This relation show that
existence the cavity without particle depend from equation of
states for particles which filled in this cavity. In the case
nonoccurrence of interaction in system we have the previous
result of Poison distribution. It is very good representative
formula. If we remember that the Laplace pressure of bubble is
$P=\frac{2\sigma}{R}$,where $R$ is the radius of void babble, we
can obtain the probability in the very know form
\begin{equation}
W(0)=exp(-\frac{8\pi\sigma R^{2}}{3kT})
\end{equation}
This is the probability formation the babble ow new phase under
first order phase transition. In general case it is the babble of
void. It is very good representation of argument to correct
description the formation cavity of void.

First from all we consider the thermodynamic properties of
systems week interacting particle. In general case we can present
the equation of state in the form
\begin{equation}
\frac{PV}{kT}=N(1-b)
\end{equation}
where $b$ is the virial coefficient. The equation of state of the
hard spheres gas can write as follows too:
\begin{equation}
PV=kTN\frac{ 1+\nu+\nu^{2}-\nu^{3}}{(1-\nu)^{3}}
\end{equation}
where $\nu=\frac{N V_{0}}{V}$ is packing factor a particle volume
is excluded, has quit different form though asymptotically is
similar to above ones. In the case of system particles which
interact through short range repulsive as hard sphere and
gravitation attractive on the far distance, in the limit of low
packing factor, equation of state can write in the form :
\begin{equation}
PV=kTN+\frac{1}{2}N^{2}\left(
Uv-\frac{W_{G}v^{\frac{1}{3}}}{2N^{\frac{1}{3} } }\right) .
\end{equation}
All equations is like a Van der Waals equations of state with the
second part of this equation representing the interaction in the
system. If we have equations of state can obtain the probability
existence the cavity without particle in the case of system
interacting particles. Mean value of size the void or cavity
without particle we can estimate compare probability to
$\frac{1}{2}$. In the case noninteracting particle we can obtain
$cV=ln 2$ or number of particle which can put up to this void is
$N=ln 2<1$. In the system of noninteracting particles can not
form the voids or cavity without particles. If we consider the
system week interaction particles we can use the equation of
states in the standard form. In this case we can estimate the
volume of void from relation 
\begin{equation}
\frac{PV}{ kT}=N(1-b)=ln 2.
\end{equation}
as $(\frac{NV_{o}}{V})\frac{V}{V_{0}}(1-b)=ln 2$  
\begin{equation}
V=V_{0}\frac{ln
2}{\nu(1-b)}
\end{equation}
where $V_{0}$ is volume of one particles, and $\nu$
as previous is packing factor. If $\nu \sim 1$ and $b\ll 1$ is
very small we can tall that in the system with week interaction
can not formed the voids. The size of voids will be the smallest
as size of particles. If the packing factor decrease and increase
the energy interaction between particle to come into the picture
the condition to formation the cavity without particle and the
size of one cavity can evaluation as 
\begin{equation}
V=V_{0}\frac{ln
2}{\nu(1-b)}
\end{equation}
As example, if $\nu\sim 0,5$b or smallest and $b=0,5$ we obtain 
that $V\sim 3V_{0}$ or bigger. Next example is the hard spheres gas. 
In this case of system particles which interact through short range 
repulsive as hard sphere we can obtain that 
\begin{equation}
V=V_{0}\frac{ln 2
(1-\nu)^{3}}{\nu(1+\nu+\nu^{2}-\nu^{3})}
\end{equation}
If $\nu\rightarrow 1$ we have that $V\rightarrow 0$. It is case 
compact packing hard sphere is obvious that in this system can not 
formed the voids. If $\nu\rightarrow 0$ as result we have that $V=V_{0}\frac{ln 2
}{\nu}$ as previous result. We can assert that the formation of
voids in the system of particles is result of interaction and
depend from concentration of particles. In the case increasing of
interaction of particles, when the relation between potential and
kinetic energy is bigger and we can not use the description of
the system in the term virial expansion we must used only
equation of state. 

In process of aggregation can produce few cellular. We can
estimate the number of cell in general case. If in process of
aggregation make up $m$ cell in the volume $\overline{V}$ which
present the volume of pattern must hold the next relation:
\begin{equation}
mV+NV_{0}=\overline{V} .
\end{equation}
From this equation can obtain that
\begin{equation}
m=\frac{\overline{V}}{V}+N\frac{V_{0}}{V}
\end{equation}
If use the previous relation between $V$ and $V_{0}$ can obtain
the number of cell as
\begin{equation}
m=N\frac{1-b_{2}}{ln 2}(1-\nu)
\end{equation}
From this equation we can conclude, that the number of cell
without particles increase if decrease the packing factor $\nu$
and decrease the interaction energy. If packing factor
$\eta\rightarrow 1$ that relation with crystal structures the
number of voids go to zero. If we consider the system non
interacting particle $b_{2}\rightarrow 0$ we obtain that
$m=N\frac{1}{ln 2}(1-\nu)$ and depended only packing factor. If
$\nu\rightarrow 1$ the number of cell came to zero too. The
number of particles in this case came to zero. If
$\eta\rightarrow 1$ the number of cellular cam to zero on other
motivation. In this case all particles formed close packing
system and in this system can not formed the cavity without
particles.

If we have the inter particle interaction energy, we can study the
thermodynamic behavior of an aggregate of such particles and describe 
the condition for the creation of new structure. The character and 
intensity of the inter particle interaction in the system of particles 
in different matter can be such that a temperature and concentration 
phase transition and produce a spatially inhomogeneous distribution 
the particles in the investigating system  \cite{lev1} \cite{lev2}. 
We must describe the first-order face transition when the external 
field present surface boundary condition on all particles. In order 
to demonstrate the mechanism and character of the phase transition 
accompanied by the formation of the inhomogeneous distribution of 
the system foreign spherical particles. For this case the free energy 
of particles in the self-consistent field approach and the many-body 
approximation can be written in the form:

\begin{equation}
F=F_{p}+F_{s}+F(n)
\end{equation}
where the first part
\begin{equation}
F_{p}=\int U(\vec{r}-\vec{r}^{\prime
})f(\vec{r})f(\vec{r}^{\prime })d\vec{r} d\vec{r}^{\prime }+...
\end{equation}
describe the free energy which cam from interaction in term function $f(\vec{r})$ 
spatially distribution particles. The simplest free energy including a interaction
terms that respect this local gauge symmetry. Se second part 
\begin{equation}
F_{s}=\int \{f(\vec{r})\ln f(\vec{r})+[1-f(\vec{r})]\ln
[1-f(\vec{r})]d\vec{r }
\end{equation}
is the entropy part free energy. This kind the entropy
part the free energy is motivation that the two classical
particle not possible occupation own space place. Next, the free
energy resulting from the coupling between distribution function
and colloidal coordinates can modelled by
\begin{equation}
F_{n}=\int f(\vec{r})
\sum_{i}W(\overrightarrow{R_{i}}-\overrightarrow{r})d\vec{r}
\end{equation}
where$W(r)$ contains microscopical information about the wetting
properties of the particles surface.

The minima of the free energy corresponds to the self-consistent
field solution for $f(r)$. Each of thermodynamic functions of
state corresponds to a solution that describes some phase of
particle arrangement. If their distribution can be inhomogeneous,
then the solution serves to find the stable phase associated with
the interaction temperature and character. The minima of the free
energy corresponds to the self-consistent field solution for
$f(r)$. Each of thermodynamic functions of state corresponds to a
solution that describes some phase of particle arrangement. If
their distribution can be inhomogeneous, then the solution serves
to find the stable phase associated with the interaction
temperature and character. If the particle solution is
disordered, then by definition the mean value $f(\vec{r})=c$,
where  $c$ is the relative particle concentration. The
concentration inhomogeneity gives rise to an additional term
$f(\vec{r})=c$ $\pm \varphi (\vec{r})$ where $ \varphi \left(
\vec{r}\right) $ is the change of the probability distribution
function of the particles. If the concentration inhomogeneities
are smooth and their scale is much longer the inter particle
distance, the quantity may be interpreted as the change of
particle composition. When passing from to continuum description,
we can write the free energy increment, associated with the
inhomogeneous particles distribution in the term of the
power series expansion in using the long-wavelength expansion of
the concentration i.e. $\varphi \left( \vec{r} ^{\prime }\right)
=\varphi \left( \vec{r}\right) +\overrightarrow{\rho }
_{i}\partial _{i}\varphi \left( \vec{r}\right) +$ $\frac{1}{2}
\overrightarrow{\rho }_{i}\overrightarrow{\rho }_{j}\partial
_{j}\partial _{i}\varphi \left( \vec{r}\right) +$. Where
$\overrightarrow{\rho }=
\overrightarrow{r}-\overrightarrow{r^{\prime }}$ distance between
two particles. In this case we may be rewrite the part free
energy, which is dependence from the change of the probability
distribution function of the particles in the form:

\begin{equation}
\triangle F\left( \varphi \right) =\int
d\overrightarrow{r}\left\{ \frac{1}{2 }l^{2}\left( \nabla \varphi
\right) ^{2}-\frac{1}{2}\mu ^{2}\varphi ^{2}+ \frac{1}{4}\lambda
\varphi ^{4}-\varepsilon \varphi \right\}
\end{equation}

Where

\begin{equation}
\mu ^{2}\equiv \left( V-\frac{kT}{c(1-c)}\right) ,\text{ }V=\int
U( \overrightarrow{\rho })d\overrightarrow{\rho },
\end{equation}
and
\begin{equation}
\text{ }l^{2}=\int U( \overrightarrow{\rho })\overrightarrow{\rho
}^{2}d\overrightarrow{\rho },
\end{equation}
$\lambda $ is the coefficient responsible for nonlinearity of system,
which is induced the many-body interaction in system particles.
The coefficient $\varepsilon =N4\pi R^{2}_{0}W $ present the
energy which include every particle through the wetting effect,
where $R_{0}$ is size of particle and $W$ present the anchoring
energy of molecules of matter with surface of particle. This
coefficient can introduce if we have the same wetting effect on
every particle. In this case in general presentation the free
energy not exist none even terms because the distribution function 
of particles satisfy the relation: $\int
f(\vec{r})d\vec{r}=N,\int \varphi (\vec{r})d\vec{r}=0.$  The
expression is the Landau free energy of a system particles which
foreign in matter, below the phase transition temperature of this
system. Thus we see that the minimum of the functional realizes a
spatially inhomogeneous macro particle distribution only provided
the sing satisfy some relation and the values of coefficients
determined by the inter particle interaction. In order to reveal
the condition under which the homogeneous particles distribution
become unstable, we have to calculate all the coefficient.
Temperature of the phase transition to new states the inhomogeneous 
distribution of particles may be determined from following relation
\begin{equation}
kT_{c}=c(1-c)V
\end{equation}
 The functional (23)is very well know functional, which description the first
-order phase transition with accompanied the cluster formation in
the system of the interacting particles. The most important contribution 
in the concentration is associated with the field configuration for 
which the value of the free energy is minimum, i.e.:
\begin{equation}
\triangle \varphi
-\frac{dV}{d\varphi }=0
\end{equation}
Where 
\begin{equation}
V=-\frac{1}{2}\mu ^{2}\varphi
^{2}+\frac{1}{4}\lambda \varphi ^{4}- \varepsilon \varphi 
\end{equation}
can be used as potential energy in our case. Substitution the solution in the
expression from the free energy yield and its variation due to
the formation of new phase. In the case when the difference of minimum 
effective potential values is greater than the barrier height, 
the solution in our case the free energy one cluster is described by 
the expression :

\begin{eqnarray}
&& \Delta F=4\pi \int\limits_{0}^{\infty }r^{2}dr\left\{
\frac{1}{2}\left( \frac{ d\varphi }{dr}\right) ^{2}+V\left(
\varphi \right) \right\}\nonumber\\
&& =-\frac{4\pi }{3} r^{3}\varepsilon +4\pi r^{2}\sigma
\end{eqnarray}
Where\ $\sigma $\ is the surface energy of the cluster boundary
that is equal the free energy corresponding to the solution of
the one-dimensional problem, i.e.
\begin{equation}
\sigma =\int\limits_{0}^{\infty }dr\left\{ \frac{1}{2}\left(
\frac{d\varphi }{dr}\right) ^{2}+V\left( \varphi \right) \right\}
=\int\limits_{0}^{\infty }d\varphi \sqrt{2V\left( \varphi \right)
}
\end{equation}
The radius of the new phase cluster by the minimum of the free energy.
It is given by $\widetilde{R}_{0}=\frac{2\sigma
}{\varepsilon }.$ As is very know we can obtain $\ \varepsilon
=\frac{2\mu \in }{\lambda ^{\frac{1}{2}}}$and $\sigma =\frac{\mu
^{3}}{3\lambda }$, then $R=\frac{\mu ^{2}l}{3\lambda
^{\frac{1}{2}}\varepsilon}$ in our case and the effective value
of the free energy variation from cluster formation is given by
\begin{equation}
\Delta F=\frac{8\pi \sigma R^{2}}{3}
\end{equation}
The probability of formation one cluster can write in the form:
\begin{equation}
P(\widetilde{R})=exp(-\frac{\Delta F}{kT})=exp(-\frac{8\pi \sigma
R^{2}}{3})
\end{equation}
They sometimes possess mixed physical properties of their
elements, but in many cases quite new properties emerge,
reflecting new structural organization of their elements. The
characteristic dimensions of the spatially inhomogeneous
distribution of the concentration of particles in the end, become
a criterion the first order phase transition with cluster
formation in the system particle. The criterion of instability
given by this relation can be interpreted as a condition for
formation of spatially non-uniform distribution at a given
temperature, which depends on the concentration of particles and
characteristic length of the new structure. If we have the some 
fixed concentration of particles in the system the process of 
aggregation finished when all particle to assemble in few cluster. 
The size of cluster will be growth to time when all particle will 
be located only cluster. Cluster formed the interaction energy 
and distribution the particle in one cluster can consider homogeneous. 
If the concentration is bigger the volume of cluster growth too, and
come the moment when the particle formed one cluster which to
hold a position on volume of all particles. Other volume will be
formed the voids. If the concentration is small we will be
observer only cluster with equilibrium size, if the concentration
is bigger as $\nu > \frac{1}{2}$ we obtain the one cluster and
few voids. In present approach we can estimate the size of voids
as characteristic length of instability the nonuniform
distribution of particle. In summary we have independently estimated 
the spontaneous formation loosely bound, order aggregates of colloidal 
particles and possible formation the cellular structures as result 
the different nature of interaction.


\begin{thebibliography}{99}                                                                                               %

\bibitem{gas}  J. Ruiz-Garsia, R. Gamez-Corrale and B. I. Ivlev,  Phys. Rev. E. {\bf 58}, 660,(1998)

\bibitem{low} H.Lowen , Phys. Rev. Lett. {\bf 74}, 1028, (1995).

\bibitem{and1}  V. G. Anderson, E. M. Terentjev, S. P. Meeker, J.Crain and
W. G. K. Poon, Eur. Phys. E.{\bf 4},11,(2001)

\bibitem{and2}  V. G. Anderson and E. M. Terentjev, Eur. Phys. E. {\bf 4}, 21,(2001)

\bibitem {chen} S. H. Chen and N. M. Amer. Phys. Rev. Lett. \textbf{51}, 2298 (1983).

\bibitem {sas} W. C. Saslaw, Gravitational physicsw of stellar and
galactic systems, Cambridge University Press, Cambridge,1987

\bibitem {nych} V. Nazarenko, A. Nych and B. Lev ,Phys.Rev.Lett ,\textbf{87},13
August (2001).

\bibitem {lev1} B. I. Lev, S. B. Chernyshuk, P. M. Tomchuk and H. Yokoyama, Phys. Rev E. \emph{ }\textbf{65},(2002).

\bibitem {lev2} B. I. Lev, H. M. Aoki and H.Y okoyama ( Phys.Rev E ,to be published)

\bibitem {huang} K. Huang, Statistical mechamics, John Wiley and
Sons, Inc., New York, 1963

\bibitem{Ric}  P. Richard, L. Oger, J.-P. Troadee and A. Gervois, Phys. Rev.
E {\bf 60}, 4551 (1999).

\bibitem {b1} D. A. Soville W. B. Russel and W. R. Schowaiter, \textit{Colloidal
Dispersions} (Cambridge University Press, Cambridge, 1989).


\end{thebibliography}
\end{document}